\documentclass{elsart}
\usepackage{epsfig}

\def\calL{{\it L}}
\def\calH{{\it H}}
\def\calF{{\it F}}
\def\BR{{\it BF}}

\begin{document}

\begin{frontmatter}
\epsfysize3cm

\hspace{-9.5cm}
\epsfbox{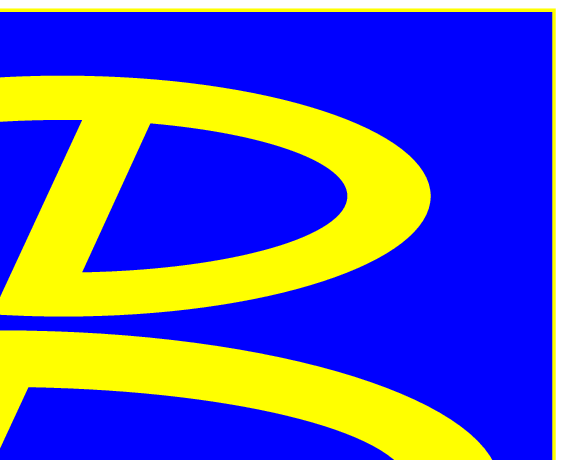}    

\vskip -3cm
\noindent
\hspace*{9.5cm}KEK Preprint 2001-63 \\
\hspace*{9.5cm}Belle Preprint 2001-11 \\

\vskip 1.5cm

\title{
Measurement of the Branching Fraction for $B\to \eta^{\prime}K$ 
and Search for $B\to \eta^{\prime}\pi^+$}

\author{The Belle Collaboration}

\begin{center}
{\normalsize
  K.~Abe$^{9}$,               
  K.~Abe$^{38}$,              
  R.~Abe$^{28}$,              
  I.~Adachi$^{9}$,            
  Byoung~Sup~Ahn$^{16}$,      
  H.~Aihara$^{40}$,           
  M.~Akatsu$^{21}$,           
  Y.~Asano$^{45}$,            
  T.~Aso$^{44}$,              
  V.~Aulchenko$^{2}$
% ...
}\end{center}

\address{
% ...
$^{2}${Budker Institute of Nuclear Physics, Novosibirsk}\\
% ...
$^{9}${High Energy Accelerator Research Organization (KEK), Tsukuba}\\
% ...
$^{16}${Korea University, Seoul}\\
% ...
$^{21}${Nagoya University, Nagoya}\\
% ...
$^{28}${Niigata University, Niigata}\\
% ...
$^{38}${Tohoku Gakuin University, Tagajo}\\
% ...
$^{40}${University of Tokyo, Tokyo}\\
% ...
$^{44}${Toyama National College of Maritime Technology, Toyama}\\
$^{45}${University of Tsukuba, Tsukuba}\\
% ...
}

\normalsize

\begin{abstract}

We report measurements for two-body charmless hadronic $B$ decays 
with an $\eta^{\prime}$ meson in the final state. 
Using $11.1\times 10^6$ $B\bar{B}$ pairs 
collected with the Belle detector,
we find $\BR(B^+\to\eta^{\prime} K^+) =
 (79^{+12}_{-11}\pm 9)\times 10^{-6}$ and
$\BR(B^0 \to\eta^{\prime} K^0) =
 (55^{+19}_{-16}\pm 8)\times 10^{-6}$, 
where the first and second errors are statistical and systematic,
respectively. 
No signal is observed in the mode $B^+\to \eta^{\prime}\pi^+$, 
and we set a 90\% confidence level upper limit of
$\BR(B^+\to\eta^{\prime} \pi^+) < 7\times 10^{-6}$. 
The $CP$ asymmetry in $B^{\pm}\to \eta^{\prime}K^{\pm}$ decays is
investigated and a limit at 90\% confidence level of $-0.20<A_{CP}<0.32$
is obtained.

\vspace{3\parskip}
\end{abstract}

\end{frontmatter}
\clearpage

Charmless hadronic $B$ decays provide a rich ground for 
studying the mechanisms of $B$ meson decay and 
the phenomenon of $CP$ violation.
The decay $B\to \eta^{\prime} K$ is an example of such a charmless decay 
with an unexpectedly large branching fraction \cite{etapke}.
Within the framework of the Standard Model, 
the $B\to \eta^{\prime} K$ decay proceeds primarily through 
$b\to s$ penguin diagrams with a contribution from the
$b\to u$ tree diagram.
Recent theory calculations \cite{theor,sanda}
underestimate the measured decay rate \cite{etapkn}
published by the CLEO collaboration. 
If the large branching fraction persists after more precise measurements,
we will likely need an additional SU(3)-singlet contribution \cite{cheng}
or new physics beyond the Standard Model to explain it.
Moreover, if the unitarity triangle angle $\phi_3$ (or $\gamma$),
defined by $\arg({{{V_{ub}^*V_{ud}}\over{-V_{cb}^*V_{cd}}}})$, is 
greater than 90 degrees,    
as suggested from interpretations of $B\to K \pi, \pi\pi$ 
results \cite{cleokpi,bellekpi} under the factorization assumption, 
$B^+\to\eta^{\prime} K^+$ will be enhanced relative 
to $B^0\to\eta^{\prime} K^0$ \cite{theor,george}. Although expected to be
small \cite{CDLU}, it is also of interest to examine the direct $CP$ asymmetry 
in $B^\pm \to \eta^{\prime} K^\pm$ decays since new physics may contribute.

In this paper we report on measurements of the branching fractions
of $B$ mesons decaying to $\eta^{\prime}K^+$, $\eta^{\prime}\pi^+$,
and $\eta^{\prime}K^0$ final states, where 
only the $K^0_S \to \pi^+ \pi^-$ transition is considered for $K^0$.
Inclusion of charge conjugate modes is implied unless
explicitly stated otherwise.
The results are obtained from data collected by
the Belle detector \cite{belle} at 
the KEKB asymmetric $e^+e^-$ storage ring \cite{accel}. 
The data sample corresponds to an integrated luminosity of
10.4 fb$^{-1}$ and consists of 11.1 million $B\bar{B}$ pairs
at the $\Upsilon(4S)$ resonance. 
The branching fractions are calculated assuming that $B^+B^-$ and 
$B^0\bar{B}^0$ are produced equally.

A detailed description of the Belle detector 
can be found in Ref. \cite{belle}; 
here we only describe briefly the parts used in this analysis. 
Charged tracks are reconstructed inside 
a 1.5 T solenoidal magnet
with a three layer double-sided silicon vertex detector (SVD)
and a central drift chamber (CDC) that consists of 
50 layers segmented into 6 axial and 5 stereo superlayers. 
The CDC covers the polar angle range between $17^\circ$ and 
$150^\circ$ in the laboratory frame and, together with the SVD, 
gives a transverse momentum resolution of 
$(\sigma_{p_t}/p_t)^2 = (0.0019 \,p_t)^2 +(0.0030)^2,$  
where $p_t$ and $\sigma_{p_t}$ are in GeV/$c$. 
Charged kaon and pion identification is performed using
a combination of three devices: 
an array  of 1188 aerogel \v{C}erenkov counters (ACC) 
covering the momentum range 1--4 GeV/$c$, 
a time-of-flight scintillation counter system (TOF) 
for track momenta below 1.5 GeV/$c$, 
and $dE/dx$ information from the CDC 
for particles with very low or high momenta. 
Situated between these devices and the solenoid coil is an
electromagnetic calorimeter (ECL) consisting of 8736 CsI(T$\ell$) crystals
with typical cross-section of 
$5.5 \times 5.5 $ cm$^2$ at the front surface and 
a depth of $16.2\,X_0$. 
The ECL provides a photon energy resolution of 
$(\sigma_E/E)^2 = 0.013^2 + (0.0007/E)^2 + (0.008/E^{1/4})^2$, where $E$ 
and $\sigma_E$ are in GeV.   

Charged tracks are required to come from the collision point and have
transverse momenta $p_t$ above 100 MeV/$c$. 
These charged tracks are then refitted with their vertex position
constrained to the run-averaged profile of $B$ meson 
decay vertices in the transverse plane. 
For $\eta^{\prime} \to \rho^0 \gamma$ decays, 
in order to reduce backgrounds, the 
minimum $p_t$ requirement is increased to 200 MeV/$c$.

Charged $K$ and $\pi$ mesons coming directly from two-body $B$ decays
are identified by combining $K/\pi$ probabilities from the CDC ($dE/dx$)
and the ACC to form a $K(\pi)$ likelihood $L_K(L_\pi)$. As these mesons
have momenta above 1.5 GeV/$c$ in the laboratory frame, TOF information
does not provide any discrimination.
Discrimination between kaons and pions is
then achieved through 
the likelihood ratio, $L_{K}$/($L_{\pi}+L_{K}$).
The performance of hadron identification is studied using 
a high momentum $D^{*+}$ data sample, where
$D^{*+}\to D^0\pi^+$, $D^0\to K^-\pi^+$. 
Selected $K$ and $\pi$ tracks are required to be 
in the same kinematic region as those from two-body $B$ decays.
We measure the pion and kaon identification efficiencies to be
$(92.4\pm 2.4)$\% and $(84.9\pm 2.1)$\%, respectively.
The rate for true pions to be misidentified as kaons is $(4.3 \pm 0.4)$\% 
while the rate for true kaons to be misidentified as pions is
$(10.4\pm 0.6)$\%.
For charged pions from $\eta^{\prime}$ decays, 
tracks identified to 
be highly kaon-like (including TOF information) are rejected. 
This loose kaon rejection requirement is studied
using $K^0_S\to \pi^+\pi^-$ events. 
The typical efficiency for charged pions is $(98.2\pm 1.0)$\%. 
$K_S^0$ candidates are reconstructed by constraining 
a pair of oppositely charged tracks with a common vertex.
This vertex is required to be distinct from 
the collision point and consistent with the $K_S^0$ flight direction.  
The invariant mass is required to be within
$\pm 30$ MeV/$c^2$ of the nominal $K_S^0$ mass. 

\begin{figure}[th] 
\begin{center}
\epsfig{file=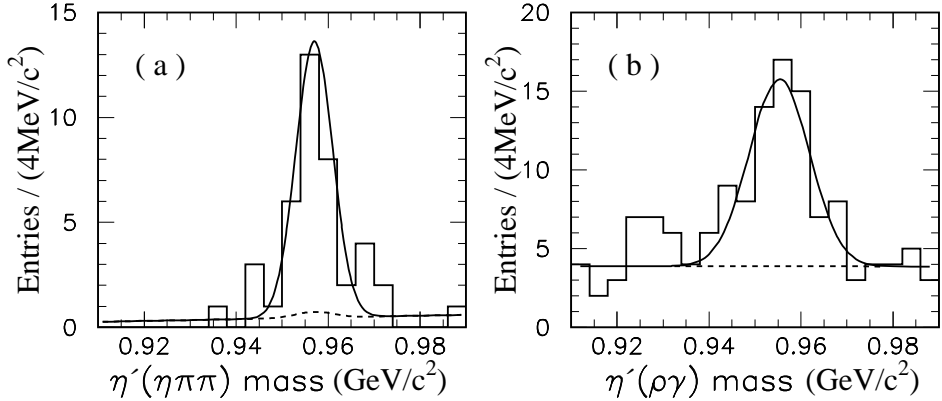,width=5.5in}
\caption{
Reconstructed mass spectra for $\eta^{\prime}\to \eta\pi^+\pi^-$ (left) 
and $\eta^{\prime}\to \rho\gamma$ (right) for
events in the $M_{bc}$-$\Delta E$ signal region  
after applying all analysis requirements.
}    
\label{fig:mass}
\end{center}
\end{figure}

Two channels have been used for $\eta^{\prime}$ reconstruction:
$\eta^{\prime} \to \eta \pi^+\pi^-, \eta \to \gamma \gamma$ and
$\eta^{\prime}\to \rho^0 \gamma$.
Candidate photons from $\eta \; (\eta^{\prime})$ decays are required
to be isolated and have energies exceeding 50 (100) MeV.  
To reject soft photon backgrounds, $\eta\to \gamma\gamma$ candidates
are selected with $|\cos\theta^*| <0.97$, where $\theta^*$ is the angle
between the photon direction in the $\eta$ rest frame 
and the $\eta$ momentum. 
The momenta of $\eta$ candidates with invariant 
mass between 500 and 570 MeV/$c^2$ are recalculated by applying an
$\eta$ mass constraint. We require $\pi^+ \pi^-$ pairs from the 
$\eta^\prime \to \rho^0 \gamma$ decay to be two oppositely charged 
tracks having an invariant mass between 550 and 920 MeV/$c^2$. 
Fig.~\ref{fig:mass} shows the mass 
distributions for $\eta^{\prime}$ candidates. 
The reconstructed mass resolutions are 12 MeV/$c^2$ for
$\eta\to \gamma\gamma$, 2.7 MeV/$c^2$ for $\eta^{\prime}\to \eta \pi^+\pi^-$,
and 8.8 MeV/$c^2$ for $\eta^{\prime}\to \rho^0\gamma$.
Candidate $\eta^{\prime}$ mesons are required to have a reconstructed mass
within $3\sigma$ of the nominal $\eta^{\prime}$ mass.

Candidate $B$ mesons are identified using the beam constrained mass
$M_{bc} =  \sqrt{E^2_{\mbox{\scriptsize beam}} - P_B^2}$ 
and the energy difference $\Delta E = E_B  - E_{\mbox{\scriptsize beam}}$, 
where $E_{\mbox{\scriptsize beam}} = 5.29$ GeV, 
and $P_{B}$ and $E_B$ are the momentum and energy of a 
$B$ candidate in the $\Upsilon(4S)$ rest frame. 
In the $B^+\to \eta^{\prime} h^+$ $(h=K,\pi)$ study, $E_B$ is computed 
with a kaon mass hypothesis for $h^+$, 
which results in a shift of +44 MeV in $\Delta E$ 
for $\eta^{\prime}\pi^+$ events. 
This shift in $\Delta E$ provides additional discrimination between the 
$\eta^{\prime} K^+$ and $\eta^{\prime} \pi^+$ final states. 
The parameterizations of the signal in $M_{bc}$ and 
$\Delta E$ are determined by a GEANT \cite{geant} based Monte Carlo (MC) 
simulation and verified using a sample of
$B^+\to \bar{D}^0\pi^+, \bar{D}^0\to K^+\pi^-\pi^0$ events. 
The Gaussian width of the signal in $M_{bc}$  is about 2.9 MeV/$c^2$, 
and mainly comes from the beam energy spread.  
The $\Delta E$ distribution is found to be slightly asymmetric 
with a small tail on the lower side due to 
$\gamma$ interactions with material in front of the calorimeter and
shower leakage out of the back side of the crystals.
This $\Delta E$ distribution is modeled by a sum of 
two Gaussians, one of which is asymmetric.
The signal region is defined as $M_{bc}>5.27$ GeV/$c^2$ and 
$-0.10$ GeV $<\Delta E<0.08$ GeV. 
Events located in the region $M_{bc}<5.265$  GeV/$c^2$
are defined as sideband events and are used for background studies.
Events with $M_{bc}>5.2$ GeV/$c^2$ and
$|\Delta E| <  250$ MeV are selected for the final analysis.  

The dominant background for two-body $B$ decay events 
comes from the $e^+e^-\rightarrow q\bar{q}$ continuum. 
In order to reduce this background, several shape variables are chosen to
distinguish spherical $B\bar{B}$ events from jet-like continuum events. 
The variable $\theta_T$ is defined as the angle between 
the candidate $\eta^{\prime}$ direction and the thrust axis 
formed by the momenta of particles not from the $B$ candidate. 
The continuum background, which accumulates mainly very 
near $|\cos\theta_T| = 1.0$,
is first reduced by requiring $|\cos\theta_T| < 0.9$. 
Two other variables used are 
$\theta_B$, the angle between the $B$ flight direction and the beam axis,
and $S_\perp$ \cite{sperp}, the scalar sum of the 
transverse momenta of all particles outside a $45^{\circ}$ cone around 
the candidate $\eta^{\prime}$ direction
divided by the scalar sum of their momenta. 
Furthermore, we introduce a set of variables
inspired from the  Fox-Wolfram moments \cite{fw}, defined as:
\begin{displaymath}
 R_l^{so} = {\sum_{i,k} |\vec{p}_i||\vec{p}_k|P_l(\cos\theta_{ik}) \over
               \sum_{i,k} |\vec{p}_i||\vec{p}_k|}
{,~~}
  R_l^{oo} = {\sum_{i,j} |\vec{p}_i||\vec{p}_j|P_l(\cos\theta_{ij}) \over
               \sum_{i,j} |\vec{p}_i||\vec{p}_j|},
\end{displaymath}
where $\vec{p}$ indicates particle momentum, 
$P_l$ is the Legendre polynomial of $l$th order, 
$k$ is either $\eta^{\prime}$ or 
$h$ from the $B$ candidate, and $i, j$ enumerate  all remaining
photons and charged particles in the event.
Since $R_1^{so}, R_3^{so},$ and $R_1^{oo}$ are found to be 
correlated with $M_{bc}$, 
we do not use them. We combine the other
five variables ($l \leq 4$) together with $\cos\theta_T$ and $S_\perp$ 
to form a Fisher discriminant $\calF$,   
\[
\calF = \sum_{l=2,4} \alpha_l R_l^{so} + \sum_{l=2,3,4} \beta_l R_l^{oo}
          + c_1 |\cos\theta_T| + c_2 S_\perp,
\]
where $\alpha_l$, $\beta_l$, $c_1$ and $c_2$ are determined by optimizing
the separation between $B\bar{B}$ events and continuum events. 
For the channel with $\rho \gamma$ in the final state,
additional discrimination is gained by using the helicity variable \calH,
which is the cosine of the angle between 
the $\pi^+$ momentum direction in the $\rho$ rest frame and 
the $\rho$ momentum direction in the $\eta^{\prime}$ rest frame.
Although there is a small non-resonant 
contribution in the $\eta^{\prime}\to \pi^+\pi^-\gamma$ process,
experimental data \cite{nonres} indicate that the helicity distribution 
can still be described by $1-\calH ^2$. 

\begin{figure}[h]
\begin{center}
\epsfig{file=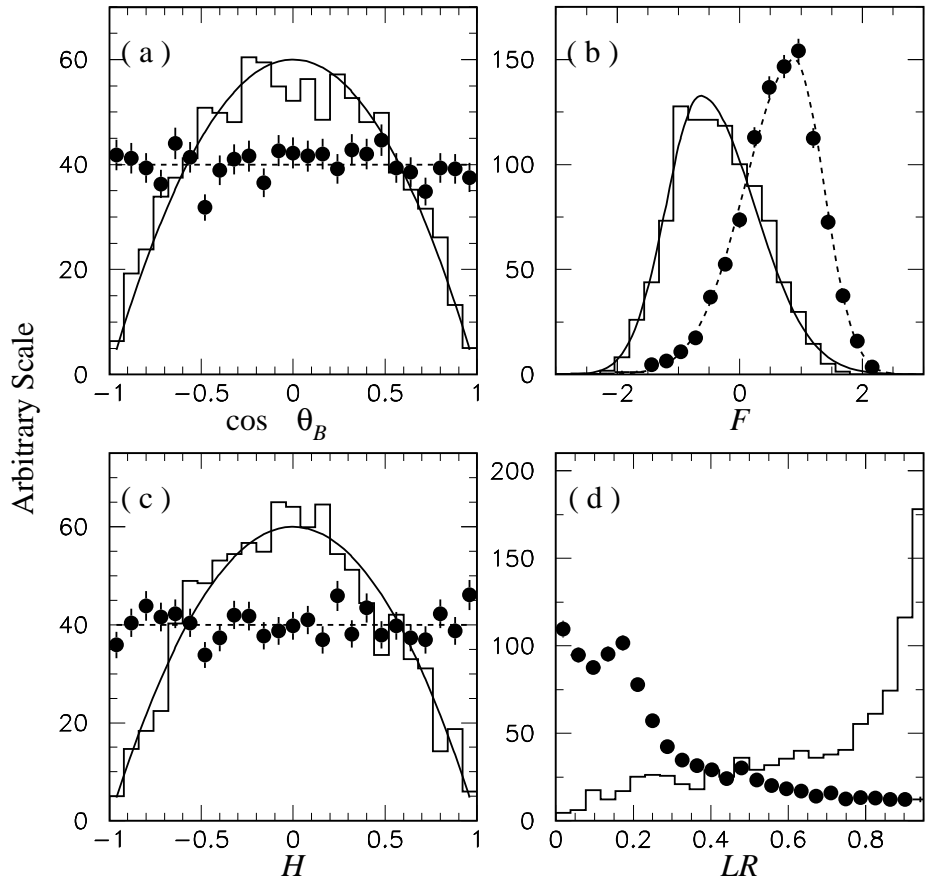,width=5.5in}
\caption{Event shape distributions for (a) $\cos\theta_B$, (b) $\calF$, 
(c) $\calH$  and (d) $LR$  for the $\rho\gamma$ mode.
Histograms represent the signal distributions (MC),
while solid points are from data outside the signal region. 
Superimposed in (a), (b) and (c) are the PDF parameterizations.
}
\label{fig:shape}
\end{center}
\end{figure}

The variables $\cos\theta_B$, $\calF$ 
and $\calH$ (for the $\rho \gamma$ channel)
are found to be independent. The probability 
density functions (PDF) for these variables are obtained 
using MC simulations for signal, and 
sideband events for $q \bar{q}$ background 
(see Fig. \ref{fig:shape}).  
These variables are then combined to form 
a likelihood ratio $LR = {\calL}_s/({\calL}_s + {\calL}_{q \bar{q}})$, 
where ${\calL}_{s (q \bar{q})}$ is 
the product of signal ($q \bar{q}$) probability densities. 
Since each channel has a different background,
we optimize the $LR$ requirement mode by mode by studying
the signal significance ($N_S/\sqrt{N_S+N_B}$) using both 
MC and data samples, where $N_S$ and $N_B$ are signal 
yields and background yields, respectively.
For instance, a loose $LR$ requirement ($LR>0.4$) for 
the $\eta\pi^+\pi^-$ mode 
keeps 83\% of the $\eta^{\prime}K^+$ signal
and reduces 74\% of the background, 
while for the $\rho\gamma$ mode a tighter requirement ($LR>0.6$) 
is used with a signal efficiency of 70\% 
and a background rejection of 88\%.
The effect of these requirements is studied by 
comparing $B^+\to \bar{D}^0\pi^+$
in data and MC with different values of $LR$.
The background from $b \to c$ transitions is negligible 
as determined by MC study.

\begin{figure}[h] 
\begin{center} 
\epsfig{file=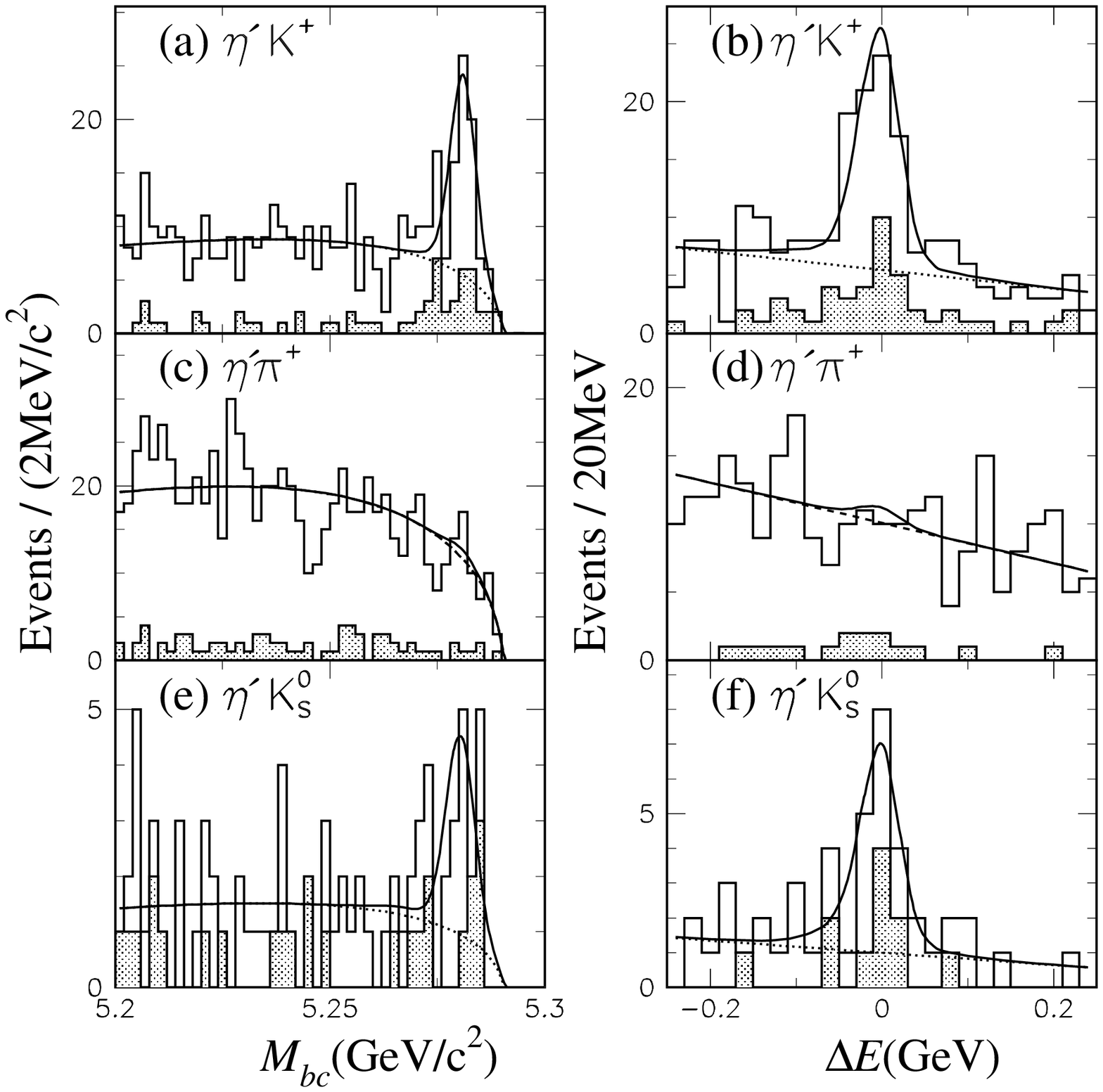,width=5.5in}    
\caption{$M_{bc}$ and $\Delta E$ projections for $\eta^{\prime}K^+$,
$\eta^{\prime}\pi^+$ and $\eta^{\prime}K^0_S$ events. The shaded histograms 
correspond to the $\eta^{\prime}\to \eta \pi^+\pi^-$ mode 
while the histograms are for all the modes combined. 
The superimposed curves show the fits to $M_{bc}$ and $\Delta E$. 
Solid curves are for signals plus backgrounds and 
dashed curves are for backgrounds only.}
\label{fig:comb}
\end{center}
\end{figure}

Fig. \ref{fig:comb}  shows the $M_{bc}$ and $\Delta E$ distributions 
for the combined $\eta\pi\pi$ and $\rho\gamma$ samples 
(and $\eta\pi\pi$ separately) in the
$\eta^{\prime}K^+, \eta^{\prime}\pi^+$ and $\eta^{\prime} K^0_S$ decay modes.  
Events in the $M_{bc}$ ($\Delta E$) plots are required to be in 
the $\Delta E$ ($M_{bc})$ signal region 
after all selection criteria are applied. 
Clear signal peaks appear in both the $\eta^{\prime}K^+$ and
$\eta^{\prime}K^0_S$ channels.
The $M_{bc}$ and $\Delta E$ distributions in each sample are 
fitted simultaneously with signal and background functions 
using an extended unbinned maximum likelihood fit.
For $N$ input candidates, the likelihood is defined as

\[
L(N_S,N_B) = \frac{e^{-(N_S+N_B)}}{N!} \prod_{i=1}^{N} 
[N_{S} P_{S_i}(M_{bc},\Delta E) + N_{B} P_{B_i}(M_{bc},\Delta E)], 
\]
where $P_{S_i}$ and $P_{B_i}$ are the 
probability densities for event $i$ to be signal 
and background for variables $M_{bc}$ and $\Delta E$, respectively.
In the extended maximum likelihood (ML) fit, the extracted yields 
for signal ($N_S$) and for
background ($N_B$) are considered separately to follow the Poisson
statistics. In this definition, the sum of $N_S$ and $N_B$ equals
the number of input candidates $N$ when the likelihood is maximized.
The $M_{bc}$ background shape is modeled with 
a smooth function \cite{argus} with parameters
determined using events 
outside of the $\Delta E$ signal region.  
The $\Delta E$ background shape is modeled by a linear function with the slope
determined from the sideband data and cross-checked using MC events.
Since significant signal peaks are 
observed in the $\eta^{\prime}K^+$ channel, 
we can expect contamination in the $\eta^\prime \pi^+$ signal region from
$\eta^{\prime}K^+$.
Therefore, an $\eta^{\prime}K^+$ signal shape is 
added in the $\Delta E$ PDF for $\eta^{\prime}\pi^+$. 
The final signal yields are then obtained by this two dimensional (2-D) fit
with the statistical significance ($\Sigma$) defined as
$\sqrt{-2\ln(\calL_0/\calL_{\rm max})}$, 
where $\calL_0$ and $\calL_{\rm max}$ denote 
the likelihood values at zero signal events 
and the best fit numbers, respectively. 
The second and third columns of Table \ref{tbl:result} list 
the fit yields and their significances. 

The systematic error for the signal yield is estimated 
by varying each parameter of the fit functions 
by $\pm 1 \sigma$ from the measured values.
The shifts in signal yield are then added in quadrature.
In order to study intrusions into the signal regions
from other rare $B$ decays,
a large MC sample of all known rare $B$ decay processes 
has been generated. No significant contributions are found.

Signal efficiencies are first obtained using MC simulations
and corrected by comparing data and MC predictions for other processes.
The  tracking efficiency is studied using 
high momentum $D$, $\eta$ and $K^*(892)$ samples. 
The $\gamma\gamma$ reconstruction efficiency is verified by measuring
the branching ratio of two $D^0$ decay channels,
$D^0\to K^-\pi^+\pi^0$ to $D^0\to K^-\pi^+$,
and four $K^*(892)$ decay channels, 
$K^+\pi^-, K^+\pi^0, K^0_S\pi^+,$ and $K^0_S\pi^0$. 
The simulation of low momentum photons is further tested by comparing 
the decay angular distribution of $\eta$ for data with MC predictions.
The systematic errors on the charged track reconstruction efficiencies of
$\eta^{\prime}\to \eta \pi^+\pi^-$ and $\eta^{\prime}\to \rho \gamma$
are estimated by comparing the ratio of 
$\eta\to \pi^+\pi^-\pi^0$ to $\eta\to \gamma\gamma$ 
in data with MC expectations. 
The reconstruction of high momentum $K_S^0$ is studied using
the ratio of $D^+\to K^0_S\pi^+$ to $D^+\to K^-\pi^+\pi^+$. 
The final systematic errors, including contributions from reconstruction 
efficiency, 
hadron identification and 2-D fits, are estimated to be 
12\% for the $\eta^{\prime}K^+$ and $\eta^{\prime}\pi^+$ modes, 
and 15\% for the $\eta^{\prime}K^0_S$ mode.   

Table \ref{tbl:result} summarizes 
the fit results for each reconstructed decay channel. 
The systematic errors for the branching fractions combine
the above systematic errors
with that from the number of $B\bar{B}$ events. 
Results from the $\eta\pi^+\pi^-$ and $\rho\gamma$ channels are then
combined by adding the $-2\ln\calL(\BR)$ functions of branching fractions
with appoximating $N = N_S + N_B$
and extracting the values and $1\sigma$ deviations at maximum $\calL(\BR)$.
Since no significant signal is seen in the $\eta^{\prime}\pi^+$ mode, 
a 90\% confidence level (C.L.) upper limit is given 
by finding the branching fraction that corresponds to
90\% of the integral of $\calL(\BR)$. 
The final upper limit is then computed by adding 
one standard deviation of the systematic error.
The final branching fractions, significance, upper limit, 
and theoretical predictions are listed in Table \ref{tbl:com}. 
With 11.1 million $B\bar{B}$ events, 
we measure the branching fractions to be
$(79^{+12}_{-11}\pm 9)\times 10^{-6}$
 for $B^+\to \eta^{\prime}K^+$ and
$(55^{+19}_{-16}\pm 8)\times 10^{-6}$
 for $B^0\to \eta^{\prime}K^0$.
The first errors are statistical and the second systematic. 
The 90\% C.L. upper limit for $B^+\to \eta^{\prime} \pi^+$ 
is $7\times 10^{-6}$.  
Our results are consistent with previous measurements given by 
 CLEO \cite{etapkn}, 
but larger than theoretical predictions \cite{theor}. 
The results are also compatible
with the preliminary results from BABAR \cite{babar}.
The upper limit on $\eta^{\prime}\pi^+$ is currently the most 
restrictive experimental result.

To study charge asymmetry, the $\eta^{\prime}K^\pm$  sample is divided into   
two subsamples: $\eta^{\prime}K^+$ and $\eta^{\prime}K^-$.
The 2-D fit in the  $M_{bc}$ {\it vs.} $\Delta E$ plane is performed for each 
subsample.
The fitted number of signal events in the $\eta^\prime\to \eta\pi\pi$, and
$\eta^\prime\to \rho\gamma$ modes
are $18.2^{+5.1}_{-4.4}$, and $13.9^{+6.0}_{-5.1}$ for $B^+$ decays, 
and $10.7^{+4.1}_{-3.4}$, and $27.9^{+7.0}_{-6.2}$ for $B^-$ decays,
respectively. 
The number of produced $B^+$ and $B^-$ are obtained 
by maximizing the product of the likelihoods for each submode.
The number of produced events is 
$398^{+90}_{-80}$ for $B^+\to\eta^{\prime}K^+$ and 
$450^{+98}_{-88}$ for $B^-\to\eta^{\prime}K^-$;
the errors are statistical only.  
The $CP$ asymmetry ($A_{CP}$) is calculated as 
$(N(B^-) - N(B^+))/(N(B^-) + N(B^+))$.  
Since the systematic errors on $\eta^{\prime}$ reconstruction and
the number of $B\bar{B}$ events cancel in the ratio, 
the systematic uncertainty of $A_{CP}$
comes mainly from the reconstruction efficiency
of charged kaons and the 2-D fit.  
The asymmetry in the $K^{\pm}$ efficiency is 
studied using inclusive charged kaons. 
The latter is measured by varying the parameters of the fit functions.
We find the systematic errors for $A_{CP}$ are
0.01 for $K^{\pm}$ reconstruction and 0.01 for the 2-D fit \cite{acpkpi}.
The $A_{CP}$ for the $B^{\pm}\to \eta^{\prime} K^{\pm}$ decay is
finally measured to be $+0.06\pm 0.15\pm 0.01$,  
which corresponds to $-0.20<A_{CP}<0.32$ at the 90\% confidence level.

In summary, we have searched for charmless hadronic $B$ decays 
with $\eta^{\prime}$ mesons in the final state.
Our results confirm that the branching fractions of
$B^+\to\eta^{\prime}K^+$ and $B^0\to \eta^{\prime}K^0$ are large. 
The branching fraction of $B^+\to\eta^{\prime}\pi^+$ is 
less than $7\times 10^{-6}$ at the 90\% C.L. 
With about 70 $B^\pm \to\eta^{\prime}K^\pm$ events, 
no significant charge asymmetry is observed.
Our value for $B \to \eta^\prime K^+$ is somewhat larger than that
for $B \to \eta^\prime K^0$, as predicted under the factorization 
assumption, but the difference is not statistically significant.
If this difference is confirmed with better precision, 
it will, along with a measurement of time dependent $CP$ asymmetry in 
$B^0\to\eta^{\prime}K^0$, provide information
to understand the underlying dynamics of $B\to \eta^{\prime} K$ decays.

We wish to thank the KEK accelerator group for excellent operations.
We acknowledge support from the Ministry of Education, Culture, Sports,
Science and Technology of Japan and
the Japan Society for the Promotion of Science;
the Australian Research Council and the Australian Department of 
Industry, Science and Resources;
the Department of Science and Technology of India;
the BK21 program of the Ministry of Education of Korea and
the Center for High Energy Physics  sponsored by the KOSEF;
the Polish State Committee for Scientific Research
under contract No.2P03B 17017;
the Ministry of Science and Technology of Russian Federation;
the National Science Council and the Ministry of Education of Taiwan;
the Japan-Taiwan Cooperative Program of the Interchange Association;
and  the U.S. Department of Energy.

 \newpage
\begin{table}
\begin{center}
\caption{Summary of results for each channel listed in the first column.
The measured signal yield ($N_S$), statistical significance ($\Sigma$), 
reconstruction efficiency ($\epsilon$), total efficiency including the secondary
branching fraction ($B_s$), and the measured branching fractions are shown.
The branching fractions are calculated by assuming that $B^+B^-$ and 
$B^0\bar{B}^0$ are produced equally from $\Upsilon(4S)$ decays. 
Uncertainties shown in 2nd and 6th columns are statistical only.
}
\label{tbl:result}
\vspace{0.1cm}
\begin{tabular}{cccccc}\hline\hline   
Mode & $N_S$ & $\Sigma$ & $\epsilon (\%)$ &$\epsilon B_s (\%)$ & $\BR(10^{-6})$\\
\hline
$\eta^{\prime}_{\eta\pi\pi}K^+$ &  $ 28.9^{+6.5}_{-5.7}$ & 9.4 & 21.7 &3.78 
  & $69^{+15}_{-14}$ \\
$\eta^{\prime}_{\rho\gamma}K^+$ & $42.5^{+9.1}_{-8.3}$ &  7.5 & 14.2 & 4.18 
 & $92^{+20}_{-18}$\\ 
$\eta^{\prime}_{\eta\pi\pi}\pi^+$ &  $0.0^{+1.2}_{-0.0}$ & 0.0 & 23.7 & 4.11
& -  \\
 $\eta^{\prime}_{\rho\gamma}\pi^+$ & $0.0^{+5.6}_{-0.0}$ & 0.0 & 15.4 & 4.55 
& - \\ 
$\eta^{\prime}_{\eta\pi\pi}K^0$  & $ 6.4^{+3.4}_{-2.7}$ &  3.5 & 20.8 &1.25
& $46^{+25}_{-20}$ \\ 
$\eta^{\prime}_{\rho\gamma}K^0$ & $10.1^{+4.4}_{-3.6}$ &  4.0 & 11.5 &1.16 
& $79^{+34}_{-28}$ \\ \hline\hline
\end{tabular}
\end{center}
\end{table}                                         

\begin{table}
\begin{center}
\caption{Combined branching fractions ($\BR$) or 90\% C.L. limit,
 significance ($\Sigma$) of Belle, CLEO \cite{etapkn}, BABAR \cite{babar}
 and theoretical expectations \cite{theor,sanda}. 
 The branching fractions are in units of $10^{-6}$.}
\label{tbl:com}
\begin{tabular}{lcrccc}\hline\hline   
Mode & This measurement($\BR$) & $\Sigma$ 
     & CLEO  & BABAR &   Theory \\\hline
$B^+\to\eta^{\prime}K^+$ & $79^{+12}_{-11}\pm 9$ & 12.0 
                         & $80^{+10}_{-9}\pm 7$  & $62\pm 18 \pm 8 $ & 21--53 \\
$B^+\to\eta^{\prime}\pi^+$ & $<7$  & 0.0 
                           & $<12$ & - & 1--3 \\
$B^0\to\eta^{\prime}K^0$ & $55^{+19}_{-16}\pm 8$ & 5.4 
                         & $89^{+18}_{-16}\pm 9$ & $< 112$ & 20--50 \\
\hline\hline
\end{tabular}
\end{center} 
\end{table}         

\newpage

 
\begin{figure}[h]
\begin{center} 
\epsfig{file=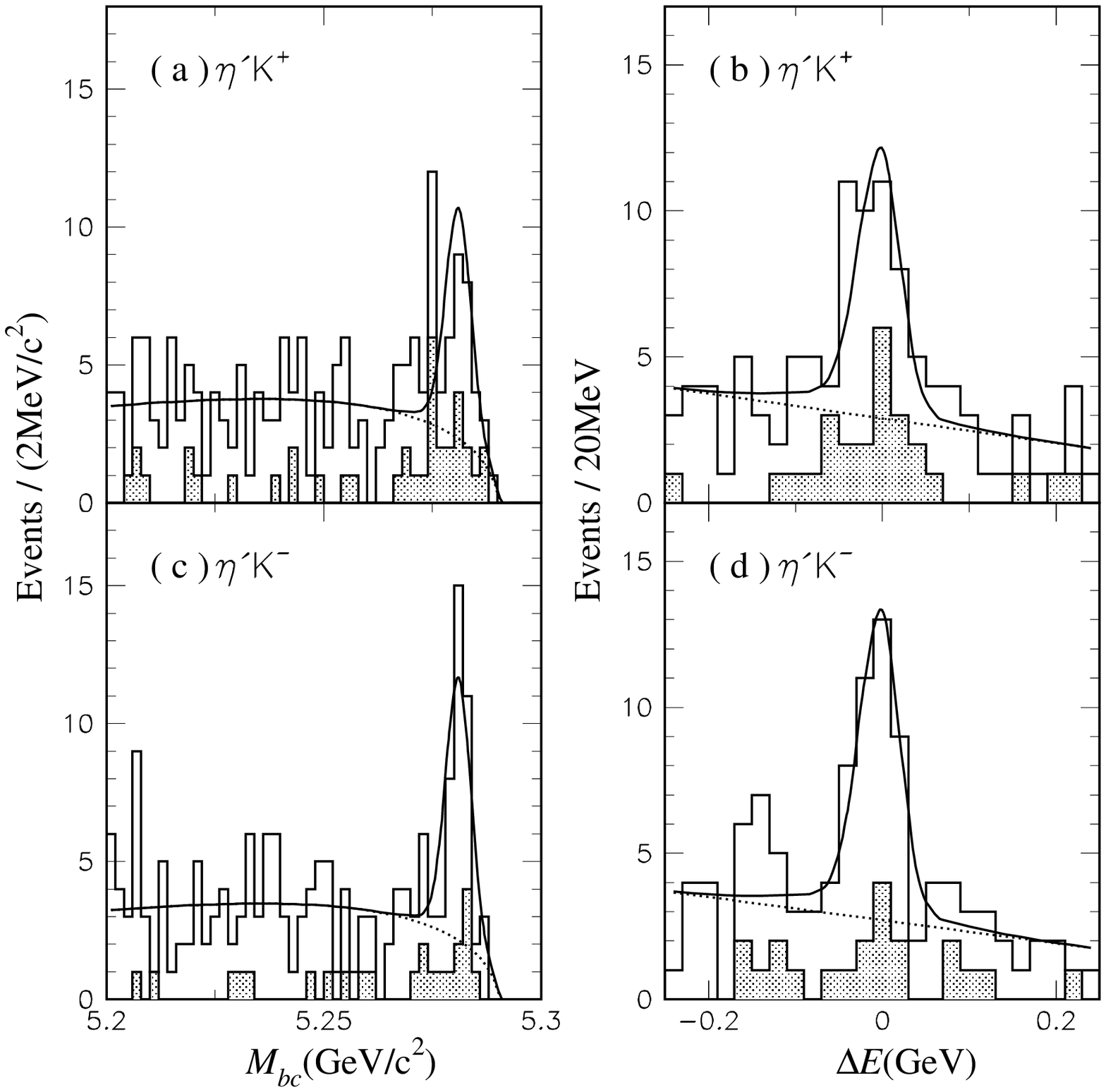,width=5.5in}    
\caption{$M_{bc}$ and $\Delta E$ projections for $\eta^{\prime}K^+$
 and $\eta^{\prime}K^-$ events. The shaded histograms
correspond to the $\eta^{\prime}\to \eta \pi^+\pi^-$ mode while the
histograms are for all the modes combined. 
The superimposed curves show the fits to $M_{bc}$ 
and $\Delta E$. Solid curves
are for signals plus backgrounds while the dashed curves 
are for backgrounds only.}
\label{fig:acp}
\end{center}
\end{figure}

\begin{thebibliography}{99}

\bibitem{etapke} CLEO Collaboration, B.H. Behrens {\it et al.},
Phys. Rev. Lett. {\bf 80} (1998) 3710.

\bibitem{theor} A. Ali, G. Kramer, and C.-D. Lu, 
Phys. Rev. D {\bf 58} (1998) 094009;
Y.-H. Chen, H.-Y. Cheng, B. Tseng, and K.-C. Yang, 
Phys. Rev. D {\bf 60} (1999) 094014; H.-Y. Cheng and K.C. Yang,
Phys. Rev. D {\bf 62} (2000) 054029.

\bibitem{sanda} E. Kou and A.I. Sanda, hep-ph/0106159 (2001). 

\bibitem{etapkn} CLEO Collaboration, S.J. Richichi {\it et al.}, 
Phys. Rev. Lett. {\bf 85} (2000) 520.

\bibitem{cheng} H.-Y. Cheng, in  Proc. XXXth Int. Conf. on High Energy Phys. 
(ICHEP2000), edited by C.S. Lim and T. Yamanaka, World Scientific, Singapore 
(2001). 

\bibitem{cleokpi} CLEO Collaboration, D. Cronin-Hennessy {\it et al.}, 
Phys. Rev. Lett. {\bf 85} (2000) 515.

\bibitem{bellekpi} Belle Collaboration, K. Abe {\it et al.},
Belle preprint 2001-5 (hep-ex/0104030), to be published in Phys. Rev. Lett.

\bibitem{george} X.-G. He, W.-S. Hou, and K.-C. Yang, 
Phys. Rev. Lett. {\bf 83} (1999) 1100.              

\bibitem{CDLU}  A. Ali, G. Kramer, and C.-D. Lu, 
Phys. Rev. D {\bf 59} (1999) 014005.              

\bibitem{belle} Belle Collaboration, A. Abashian {\it et al.}, 
KEK Progress Report 2000-4 (2000), 
to  be published in Nucl. Inst. and Meth. A. 

\bibitem{accel} KEK accelerator group, KEKB B-Factory Design Report, 
KEK Report 95-7 (1995), unpublished.

\bibitem{geant} R. Brun {\it et al.}, GEANT 3.21, 
CERN Report No. DD/EE/84-1 (1987).

\bibitem{sperp} CLEO Collaboration, R. Ammar {\it et al.},
 Phys. Rev. Lett. {\bf 71} (1993) 674.

\bibitem{fw} G. Fox and S. Wolfram, 
Phys. Rev. Lett. {\bf 41} (1978) 1581.

\bibitem{nonres} S.I. Bityukov {\it et al.}, 
Z. Phys. C {\bf 50} (1991) 451; 
Crystal Barrel Collaboration, A.~Abele {\it et al.}, 
Phys. Lett. B {\bf 402} (1997) 195.

\bibitem{argus} ARGUS Collaboration, H. Albrecht {\it et al.}, 
Phys. Lett. B {\bf 241} (1990) 278.

\bibitem{babar} BABAR Collaboration, T. Champion, 
in Proc. XXXth Int. Conf. on High Energy Phys. (ICHEP2000),
edited by C.S. Lim and T. Yamanaka, 
World Scientific, Singapore (2001).

\bibitem{acpkpi} For a more detailed description of 
the systematic errors in $A_{CP}$ from $K^{\pm}$ 
reconstruction, see Belle Collaboration, K. Abe {\it et al.},
Belle preprint 2001-9 (hep-ex/0106095), to be published in Phys. Rev. D. 

\end{thebibliography}
\end{document}